\documentclass [11pt,a4paper] {article}

\usepackage[cp1252]{inputenc}

\usepackage{amssymb}
\usepackage{amsmath}
\usepackage{amsfonts,amssymb}

\usepackage[dvips]{graphicx}

\usepackage{bbm}

\DeclareMathAlphabet{\mathpzc}{OT1}{pzc}{m}{it}

\newcommand*\rfrac[2]{{}^{#1}\!\!/\!_{#2}}

\begin{document}

\title{Quantum Supervaluationism}

\author{Arkady Bolotin\footnote{$Email: arkadyv@bgu.ac.il$\vspace{5pt}} \\ \textit{Ben-Gurion University of the Negev, Beersheba (Israel)}}

\maketitle

\begin{abstract}\noindent As it is known, neither classical logical conjunction ``\textit{and}'' nor classical logical alternative ``\textit{either…or}'' can replace ``+'' representing a linear superposition of two quantum states. Therefore, to provide a logical account of the quantum superposition, one must either reconsider the standard interpretation of quantum mechanics (making it fit for classical bivalent logic) or replace the standard logic with a deviant logic suitable for describing the superposition. In the paper, a supervaluation approach to the description of the quantum superposition is considered. In accordance with this approach, the indefinite propositions, which correspond to the superposition states, lack truth-values of any kind even granting that their compounds (such as logical alternative ``\textit{either…or}'') can have truth-values. As an illustration, the supervaluationist account of the superposition of spin states is presented.\\

\end{abstract}

\section{Introduction: Treating vagueness in quantum mechanics}  

\noindent The application of the laws of classical logic where it combines the bivalent truth values – i.e., truth and falsity – of individual propositions to the principle of quantum superposition appears to lead to a problem.\\

\noindent To be sure, let us consider a typical Stern-Gerlach experiment in which a non-uniform magnetic field splits a beam of silver atoms being sent through it into two beams depending on possible spin angular momentum values $+\rfrac{\hbar}{2}$ and $-\rfrac{\hbar}{2}$ along a selected axis $j \in \{x,y,z\}$ (denoted respectively as “$j$-up” and “$j$-down”) of each atom.\\

\noindent Suppose that in the sequential experiments linking multiple Stern-Gerlach apparatuses, after the first experiment the beam ``$z$-down'' is blocked at the same time as the beam ``$z$-up'' is directed to the second experiment measuring the spin state of the atoms along the $x$-axis. According to the orthodox quantum theory, before the second experiment ends with some result, the spin state of the particle $|{\Psi}_{z+}\rangle$ is described by the superposition of the spin states $|{\Psi}_{x+}\rangle$ and $|{\Psi}_{x-}\rangle$ pointing in “$x$-up” and “$x$-down” directions, namely,\smallskip

\begin{equation} \label{1} 
      |{\Psi}_{z+}\rangle
      =
      c_1 |{\Psi}_{x+}\rangle + c_2 |{\Psi}_{x-}\rangle
      \;\;\;\;  ,
\end{equation}
\smallskip

\noindent where $c_1$ and $c_2$ are superposition coefficients (satisfying the condition $|c_1|^2+|c_2|^2=1$).\\

\noindent Let $Z_{+}$ denote the proposition that the particle exists in the spin state ``$z$-up'' till the second experiment produces the result. Furthermore, let $X_{+}$ stand for the proposition that at the same time the particle is in the spin state ``$x$-up'', whereas $X_{-}$ signify the alternative proposition that this particle is in the spin state ``$x$-down''.\\

\noindent As follows, in the situation described by the superposition (\ref{1}), the proposition $Z_{+}$ is definite and has a value of true at the same time that using a vague semantics \cite{Williamson} the propositions $X_{+}$ and $X_{-}$ can be described as \textit{indefinite}.\\

\noindent Now, let us try to describe the superposition (\ref{1}) within the framework of classical logic.\\

\noindent If the superposition of the spin states $|{\Psi}_{x+}\rangle$ and $|{\Psi}_{x-}\rangle$ were to be represented by classical conjunction ``\textit{and}'' on the values of two indefinite propositions $X_{+}$ and $ X_{-}$, then one would get\smallskip

\begin{equation} \label{2} 
      Z_{+}
      =
      X_{+} \land X_{-}
      \;\;\;\;  ,
\end{equation}
\smallskip

\noindent which would imply that until the result of the second experiment is known, the particle is in the state of being both spin-up and spin-down (i.e., simultaneously in two beams spatially separated along the $x$-axis) which would drastically contradict our experience of the everyday world.
$\,$\footnote{\label{f1}Admittedly, the formula (\ref{2}) considered in conjunction with the hypothesis that the quantum superposition principle governs all the natural processes implies that if the initial state is a particle's state of being both spin-up and spin-down, then the final state must be an apparatus's state of being simultaneously in two macroscopically and perceptively different states. In this way, the formula (\ref{2}) gives rise to \textit{the measurement (or macro-objectification) problem} considered by at least some authors as a serious physical problem (see, for example, \cite{Gisin}).\vspace{5pt}}\\

\noindent However, if the superposition of the states $|{\Psi}_{x+}\rangle$ and $|{\Psi}_{x-}\rangle$ were to be represented by classical exclusive disjunction ``\textit{either…or}'' on the values of $X_{+}$ and $ X_{-}$, then the macro-objectification problem would be avoided, namely,\smallskip

\begin{equation} \label{3} 
      Z_{+}
      =
      X_{+} \,\underline{\lor}\, X_{-}
      =
      (X_{+} \lor X_{-}) \land \neg (X_{+} \land X_{-})
      \;\;\;\;  ,
\end{equation}
\smallskip

\noindent but it would suggest that two components of the particle's spin could be simultaneously known (the spin $+\rfrac{\hbar}{2}$ along the $z$-axis and, for example,  the spin $+\rfrac{\hbar}{2}$ along the $x$-axis), which would contradict the quantum mechanical impossibility of simultaneous measurement of conjugate pairs of properties (namely, spin measurements along different axes).\\

\noindent Thus, neither of the equations (\ref{2}) and (\ref{3}) is right.\\

\noindent Granted the proposition $X_{-}$ can be considered as the negation of the proposition $X_{+}$, this means that \textit{the principle of excluded middle} (stating that for every bivaluation of $X_{+}$ and $\neg X_{+}$ the disjunction $X_{+} \lor \neg X_{+}$ must be true) does not hold in the case of the superposition (\ref{1}). Hence, classical logic and the quantum superposition principle are incompatible.\\

\noindent Such an incompatibility suggests that in order to provide a logical account of the superposition (\ref{1}), one should either reconsider the standard formulation and interpretation of quantum mechanics making it consistent with classical logic or replace the standard logic with a deviant logic suitable for describing the superposition (\ref{1}).\\

\noindent Let us dwell on the second option
$\,$\footnote{\label{f2}A concise yet thorough analysis of the first option (commonly known as \textit{the problem of an interpretation of quantum mechanics}) can be found, for instance, in \cite{Tammaro}.\vspace{5pt}} and consider how this revision of logic can be executed. Among various (possibly uncountable) ways to implement this revision, the two most obvious choices are \textit{proliferation of truth-values} (i.e., \textit{logical many-valuedness} also known as \textit{quantum logic}) and \textit{supervaluationism}.\\

\noindent On the approach of logical many-valuedness, the indefinite propositions $X_{+}$ and $\neg X_{+}$ are assigned truth-values that lie between 1 (which stands for \textit{the truth}) and 0 (which stands for \textit{the falsehood}). In some works, a system of many-valued logic with the 3-valued set of truth-values is favored (see \cite{Putnam}), while the others (such as papers \cite{Pykacz94, Pykacz95, Pykacz00, Pykacz10, Pykacz11, Pykacz15} and book \cite{Pykacz15b}) insist on a system of logic with an infinite number of truth-values.\\

\noindent As to the basic insight of supervaluationism, it boils down to the following two thoughts: First, a vague language is one that allows several \textit{precisifications} (i.e., ways of making a statement more precise); second, when a language allows several precisifications, its semantics is fixed only insofar as -- and exactly insofar as -- all those precisifications agree \cite{Varzi}. In particular, this means that a compound of simple statements is true if it is \textit{super-true}, that is, true on every admissible precisification, and it is false if it is \textit{super-false}, that is, false on every admissible precisification, even supposing the simple statements lack a truth-value.\\

\noindent One can infer from here that using a supervaluationary semantics the disjunction $X_{+} \!\lor \neg X_{+}$ can be described as super-true, i.e., true in all admissible states of the particle, and the conjunction $X_{+} \!\land \neg X_{+}$ as super-false, i.e., false in all admissible states of the particle, regardless of whether or not the constituent propositions $X_{+}$ and $\neg X_{+}$ have a truth value.\\

\noindent Clearly, this can neatly fit the impossibility of knowing a particle's spin along one axis if it is already known along the other. Besides, while admitting ``truth-value gaps'' the supervaluationist approach allows one to retain such principles of standard logic as the principles of excluded middle and non contradiction \cite{Keefe}.\\

\noindent Given that the supervaluationist approach is virtually unexplored in quantum mechanics, let us examine it a bit further in this paper.\\

\section{Preliminaries: Truth values of quantum phenomena}  

\noindent Let us consider a lattice $L(\mathcal{H})$ of closed subspaces of a (separable) Hilbert space $\mathcal{H}$, where the partial order $\le$ corresponds to the subset relation $\subseteq$ between the subspaces of $\mathcal{H}$, the lattice meet $\sqcap$ corresponds to the intersection $\cap$ of the said subspaces and the lattice join $\sqcup$ is the closed span of their union $\cup$. The lattice $L(\mathcal{H})$ is bounded with the trivial space $\{0\}$ as the bottom and the identical subspace $\mathcal{H}$ as the top.\\

\noindent As any closed subspace of $\mathcal{H}$ is the column space (range) of some unique projection operator on $\mathcal{H}$, there is a one-to-one correspondence between such subspaces and the projection operators. Hence, one can take projection operators to be elements of $L(\mathcal{H})$.\\

\noindent Specifically, let a set $\mathcal{O}$ of commutative (nontrivial) projection operators $\hat{P}_A, \hat{P}_B, \dots$ on $\mathcal{H}$ representing simultaneously verifiable propositions $A, B, \dots$ that are related to properties of a physical system be called \textit{a context} $\mathcal{O}$. Consider the sublattice $L_\mathcal{O}$ of $L(\mathcal{H})$ generated by the column and null spaces of the projection operators in the context $\mathcal{O}$. Let the subset relation between the column space and the null space of any two projection operators $\hat{P}_A$ and $\hat{P}_B$ in the context $\mathcal{O}$ be as follows\smallskip

\begin{equation} \label{4} 
   \mathrm{ran}(\hat{P}_A) 
   \subseteq
   \mathrm{ker}(\hat{P}_B)
   =
   \mathrm{ran}(\hat{1} - \hat{P}_B) 
   \;\;\;\;  ,
\end{equation}
\smallskip

\noindent where $\hat{1}$ denotes the identity operator on $\mathcal{H}$ whose column space is equal to the identical subspace $\mathcal{H}$, namely, $\mathrm{ran}(\hat{1})=\mathcal{H}$. Then, one can impose the algebraic structure $(\mathcal{O},\sqcap,\sqcup)$ consisting of the context $\mathcal{O}$ and two binary operations $\sqcap$ and $\sqcup$ upon $L_\mathcal{O}$ within which the partial order $\hat{P}_A \le (\hat{1} - \hat{P}_B)$ is defined if $\hat{P}_A \sqcap (\hat{1} - \hat{P}_B) = \hat{P}_A (\hat{1} - \hat{P}_B) = \hat{P}_A$ or $\hat{P}_A \sqcup (\hat{1} - \hat{P}_B) = \hat{P}_A + (\hat{1} - \hat{P}_B) - \hat{P}_A (\hat{1} - \hat{P}_B) = (\hat{1} - \hat{P}_B)$. In this way, the meet and join of the projection operators $\hat{P}_A$ and $\hat{P}_B$ in the context $\mathcal{O}$ are given by\smallskip

\begin{equation} \label{5} 
    \hat{P}_A \sqcap \hat{P}_B
    =
    \hat{P}_A \hat{P}_B
    =
    \hat{0}
   \;\;\;\;  ,
\end{equation}

\begin{equation} \label{6} 
    \hat{P}_A \sqcup \hat{P}_B
    =
    \hat{P}_A + \hat{P}_B
   \;\;\;\;  ,
\end{equation}
\smallskip

\noindent where $\hat{0}$ denotes the null operator on $\mathcal{H}$ whose column space is the trivial space $\{0\}$, i.e., $\mathrm{ran}(\hat{0})=\{0\}$.\\

\noindent Let us consider the truth-value assignment of the projection operators in the context $\mathcal{O}$.\\

\noindent Let $\S^{\mathcal{V}}_{N}(\diamond) = {[\![ \,\diamond\, ]\!]}_v$, where the symbol $\diamond$ can be replaced by any proposition (compound or simple) associated with the context $\mathcal{O}$, refer to \textit{a valuation}, that is, a mapping from a set of such propositions denoted by $S_{\mathcal{O}}=\{\diamond\}$ to a set $\mathcal{V}_N = \{ \mathfrak{v} \}$ where $\mathfrak{v}$ are truth-values ranging from 0 to 1 while $N$ is the cardinality of the set $\{ \mathfrak{v} \}$:\smallskip

\begin{equation} \label{7} 
   \S^{\mathcal{V}}_{N}
   :
   S_{\mathcal{O}} \to \mathcal{V}_N
   \;\;\;\;  .
\end{equation}
\smallskip

\noindent At the same time, assume that there is a homomorphism $f:\mathcal{O} \to S_{\mathcal{O}}$ such that there is a truth-value assignment function $v$ that maps an element $\hat{P}_{\diamond}$ in $\mathcal{O}$ to the truth value of the corresponding proposition $\diamond$ in $S_{\mathcal{O}}$, explicitly,\smallskip

\begin{equation} \label{8} 
   v(\hat{P}_{\diamond}) 
   = 
   {[\![ \,\diamond\, ]\!]}_v
   \;\;\;\;  .
\end{equation}
\smallskip

\noindent In accordance with this assumption, the valuation of conjunction, disjunction and negation on the values of the propositions associated with the context $\mathcal{O}$ can be decided through the following axioms\smallskip

\begin{equation} \label{9} 
    v(\hat{P}_A \hat{P}_B)
    =
   {[\![ A \!\land\! B ]\!]}_v
   \;\;\;\;  ,
\end{equation}

\begin{equation} \label{10} 
    v(\hat{P}_A + \hat{P}_B)
    =
   {[\![ A \!\lor\! B ]\!]}_v
   \;\;\;\;  ,
\end{equation}

\begin{equation} \label{11} 
    v(\hat{1}-\hat{P}_{\diamond})
    =
   {[\![ \neg \,\diamond\, ]\!]}_v
   \;\;\;\;  .
\end{equation}
\smallskip

\noindent Suppose that the system is prepared in a pure normalized state $|{\Psi}_A\rangle$ that lies in the column space of some projection operator $\hat{P}_A$ in the context $\mathcal{O}$. Then, being in the state $|{\Psi}_A\rangle$ is subject to the assumption that the truth-value assignment function $v$ must assign the truth-value 1 to the operator $\hat{P}_A$ and, in this way, the proposition $A$, namely, $v(\hat{P}_A) = {[\![ A  ]\!]}_v =1$. Conversely, if $v(\hat{P}_A) = {[\![ A  ]\!]}_v =1$, then it is reasonable to assume that the system is prepared in the state $|{\Psi}_A\rangle$ that lies in $\mathrm{ran}(\hat{P}_A)$. Accordingly, these two assumptions can be written down as the following logical biconditional\smallskip

\begin{equation} \label{12} 
   |{\Psi}_A\rangle
   \in
   \mathrm{ran}(\hat{P}_A)
   \iff
   v(\hat{P}_A)
   =
   {[\![ A ]\!]}_v
   =
   1
   \;\;\;\;  .
\end{equation}
\smallskip

\noindent On the other hand, in view of orthogonality of all the projection operators in $\mathcal{O}$, the prepared vector $|{\Psi}_A\rangle$ must be also in the null space of any other operator in the context $\mathcal{O}$, say, $\hat{P}_B$, and the function $v$ must assign the truth value 0 to $\hat{P}_B$, namely, $v(\hat{P}_B) = {[\![ B  ]\!]}_v = 0$. Hence, by the same token,\smallskip

\begin{equation} \label{13} 
   |{\Psi}_A\rangle
   \in
   \mathrm{ker}(\hat{P}_B)
   \iff
   v(\hat{P}_B)
   =
   {[\![ B ]\!]}_v
   =
   0
   \;\;\;\;  .
\end{equation}
\smallskip

\noindent Allowing for the fact that the column space of any projection operator does not equal to its null space, it must be that $|{\Psi}_A\rangle \notin \mathrm{ker}(\hat{P}_A)$ if $|{\Psi}_A\rangle \in \mathrm{ran}(\hat{P}_A)$ as well as $|{\Psi}_A\rangle \notin \mathrm{ran}(\hat{P}_B)$ if $|{\Psi}_A\rangle \in \mathrm{ker}(\hat{P}_B)$. Thus, the definiteness of the propositions associated with the selected (by the preparation of the system in the state $|{\Psi}_A\rangle$) context $\mathcal{O}$ can be written down as the bivaluation, e.g., ${[\![ A ]\!]}_v \neq 0$ if ${[\![ A ]\!]}_v = 1$ and ${[\![ B ]\!]}_v \neq 1$ if ${[\![ B ]\!]}_v = 0$.\\

\noindent Furthermore, in the selected context $\mathcal{O}$, the truth values of conjunction, disjunction and negation can be expressed with the basic operations of arithmetic or by the minimum and maximum functions, specifically,\smallskip

\begin{equation} \label{14} 
   |\Psi_A\rangle
   \in
   \left\{
      \begin{array}{l}
         \mathrm{ran}(\hat{P}_A)\\
         \mathrm{ker}(\hat{P}_B)
      \end{array}
   \right.
   \iff
   \left\{
      \begin{array}{l}
         {[\![ A \!\land\! B ]\!]}_v = \min\left\{{[\![ A ]\!]}_v, {[\![ B ]\!]}_v\right\} = 0\\
         {[\![ A \!\lor\! B ]\!]}_v   = \max\left\{{[\![ A ]\!]}_v, {[\![ B ]\!]}_v\right\} = 1\\
         {[\![ \neg \,\diamond\, ]\!]}_v = 1 - {[\![ \,\diamond\, ]\!]}_v
      \end{array}
   \right.
   \;\;\;\;  .
\end{equation}
\smallskip

\noindent Now, by contrast, suppose that the system is prepared in a pure state $|\Psi\rangle$ that does not lie in the column or null space of any projection operator $\hat{P}_{\diamond}$ of the context $\mathcal{O}$, i.e., $|\Psi\rangle \notin \mathrm{ran}(\hat{P}_{\diamond})$ and $|\Psi\rangle \notin \mathrm{ker}(\hat{P}_{\diamond})$ (which can happen if $|\Psi\rangle$ is arranged in the column space of some projection operator belonging to a different from $\mathcal{O}$ context whose elements do not commute with ones of $\mathcal{O}$). Under the valuation assumptions (\ref{12}) and (\ref{13}), the truth-value function $v$ must assign neither 1 nor 0 to any operator $\hat{P}_{\diamond} \in \mathcal{O}$, namely,\smallskip

\begin{equation} \label{15} 
   |\Psi\rangle
   \notin
   \left\{
      \begin{array}{l}
         \mathrm{ran}(\hat{P}_{\diamond})\\
         \mathrm{ker}(\hat{P}_{\diamond})
      \end{array}
   \right.
   \iff
   \left\{
      \begin{array}{l}
         v(\hat{P}_{\diamond}) \neq 1\\
         v(\hat{P}_{\diamond}) \neq 0
      \end{array}
   \right.
   \;\;\;\;  .
\end{equation}
\smallskip

\noindent This means that the set of the propositions $S_{\mathcal{O}}=\{\diamond\}$ associated with the nonselected context $\mathcal{O}$ cannot be bivalent under $v$, i.e., ${[\![ \,\diamond\, ]\!]}_v \notin \mathcal{V}_2$.\\

\noindent As stated by the approach of logical infinite-valuedness, the failure of bivalence in the nonselected context $\mathcal{O}$ requires that the operator $\hat{P}_{\diamond} \in \mathcal{O}$ should be assigned the truth value $\mathfrak{v} \in \mathcal{V}_{\infty}$ that lies between 1 and 0. Accordingly, in an arbitrary state $|\Phi\rangle$ one should have\smallskip

\begin{equation} \label{16} 
   v(\hat{P}_{\diamond})
   =
   \langle{\Phi}|\hat{P}_{\diamond}|{\Phi}\rangle
   \in
   \{
      \mathfrak{v} \in \mathbb{R}
      \,
      |
      \,
      0 \le \mathfrak{v} \le 1
   \}
   \;\;\;\;  ,
\end{equation}
\smallskip

\noindent where the value $\langle{\Phi}|\hat{P}_{\diamond}|{\Phi}\rangle$ represents the degree to which the proposition $\diamond$ is true in the state $|\Phi\rangle$.\\

\noindent But the failure of bivalence might also imply another approach according to which any element of the nonselected context $\mathcal{O}$ carries no truth value at all, specifically,\smallskip

\begin{equation} \label{17} 
   |\Psi\rangle
   \notin
   \left\{
      \begin{array}{l}
         \mathrm{ran}(\hat{P}_{\diamond})\\
         \mathrm{ker}(\hat{P}_{\diamond})
      \end{array}
   \right.
   \iff
   \left\{
      v(\hat{P}_{\diamond})
   \right\}
   =
   \left\{
      {[\![ \,\diamond\, ]\!]}_v
   \right\}
   =
   \emptyset
   \;\;\;\;  .
\end{equation}
\smallskip

\noindent Within the said approach, the operators $\hat{1}$ and $\hat{0}$ can be treated as \textit{super-true} and \textit{super-false} since under the truth-value assignment function $v$ they are definite in any arbitrary state $|\Phi\rangle \in \mathcal{H}$, that is,\smallskip

\begin{equation} \label{18} 
   |\Phi\rangle
   \in
   \left\{
      \begin{array}{l}
         \mathrm{ran}(\hat{1}) = \mathcal{H}\\
         \mathrm{ker}(\hat{0}) = \mathcal{H}
      \end{array}
   \right.
   \iff
   \left\{
      \begin{array}{l}
         v(\hat{1}) = 1\\
         v(\hat{0}) = 0
      \end{array}
   \right.
   \;\;\;\;  .
\end{equation}
\smallskip

\noindent In this way, while equating $\hat{1}$ and $\hat{0}$ with the ``\textit{super-truth}'' and the ``\textit{super-falsity}'', the approach based on the assumption (\ref{17}) will result in a non-bivalent logic with the propositions of the nonselected context giving rise to truth-value gaps. Accordingly, this approach can be called ``\textit{quantum supervaluationism}''.\\

\section{The supervaluationist account of the quantum superposition}  

\noindent Consider the complex Hilbert space $\mathcal{H} = \mathbb{C}^2$ formed by complex $2 \times 2$ matrices related to the states for the spin of a spin-$\rfrac{1}{2}$ particle.\\

\noindent Provided $|{\Psi}_{z\pm}\rangle$ and $|{\Psi}_{x\pm}\rangle$ are the normalized eigenvectors of the corresponding Pauli spin matrices, the projection operators $\hat{P}_{z\pm}$ and $\hat{P}_{x\pm}$ on $\mathbb{C}^2$ are given explicitly by\smallskip

\begin{equation} \label{19} 
   \hat{P}_{z+}
   =
   \left[
      \!\!
      \begin{array}{r r}
         1 & 0\\
         0 & 0
      \end{array}
      \!\!
   \right]
   \;
   ,
   \;
   \hat{P}_{z-}
   =
   \left[
      \!\!
      \begin{array}{r r}
         0 & 0\\
         0 & 1
      \end{array}
      \!\!
   \right]
   \;\;\;\;  ,
\end{equation}

\begin{equation} \label{20} 
   \hat{P}_{x+}
   =
   \frac{1}{2}
   \left[
      \!\!
      \begin{array}{r r}
         1 & 1\\
         1 & 1
      \end{array}
      \!\!
   \right]
   \;
   ,
   \;
   \hat{P}_{x-}
   =
   \frac{1}{2}
   \left[
      \!\!
      \begin{array}{r r}
         1 & -1\\
         -1 & 1
      \end{array}
      \!\!
   \right]
   \;\;\;\;  ,
\end{equation}
\smallskip

\noindent at the same time as their column and null spaces are\smallskip

\begin{equation} \label{21} 
   \mathrm{ran}(\hat{P}_{z+})
   =
   \bigg\{
   \left[
      \!\!
      \begin{array}{r}
         a\\
         0
      \end{array}
      \!\!
   \right]
   \bigg{|}\,
   a \in \mathbb{R}
   \bigg\}
   \;
   ,
   \;
   \mathrm{ker}(\hat{P}_{z+})
   =
   \bigg\{
   \left[
      \!\!
      \begin{array}{r}
         0\\
         a
      \end{array}
      \!\!
   \right]
   \bigg{|}\,
   a \in \mathbb{R}
   \bigg\}
   \;\;\;\;  ,
\end{equation}

\begin{equation} \label{22} 
   \mathrm{ran}(\hat{P}_{z-})
   =
   \bigg\{
   \left[
      \!\!
      \begin{array}{r}
         0\\
         a
      \end{array}
      \!\!
   \right]
   \bigg{|}\,
   a \in \mathbb{R}
   \bigg\}
   \;
   ,
   \;
   \mathrm{ker}(\hat{P}_{z-})
   =
   \bigg\{
   \left[
      \!\!
      \begin{array}{r}
         a\\
         0
      \end{array}
      \!\!
   \right]
   \bigg{|}\,
   a \in \mathbb{R}
   \bigg\}
   \;\;\;\;  ,
\end{equation}

\begin{equation} \label{23} 
   \mathrm{ran}(\hat{P}_{x+})
   =
   \bigg\{
    \left[
      \!\!
      \begin{array}{r}
         a\\
         a
      \end{array}
      \!\!
   \right]
   \bigg{|}\,
   a \in \mathbb{R}
   \bigg\}
   \;
   ,
   \;
   \mathrm{ker}(\hat{P}_{x+})
   =
   \bigg\{
    \left[
      \!\!
      \begin{array}{r}
         a\\
        -a
      \end{array}
      \!\!
   \right]
   \bigg{|}\,
   a \in \mathbb{R}
   \bigg\}
   \;\;\;\;  ,
\end{equation}

\begin{equation} \label{24} 
   \mathrm{ran}(\hat{P}_{x-})
   =
   \bigg\{
   \left[
      \!\!
      \begin{array}{r}
          a\\
         -a
      \end{array}
      \!\!
   \right]
   \bigg{|}\,
   a \in \mathbb{R}
   \bigg\}
   \;
   ,
   \;
   \mathrm{ker}(\hat{P}_{x-})
   =
   \bigg\{
   \left[
      \!\!
      \begin{array}{r}
          a\\
          a
      \end{array}
      \!\!
   \right]
   \bigg{|}\,
   a \in \mathbb{R}
   \bigg\}
   \;\;\;\;  .
\end{equation}
\smallskip

\noindent As it can be readily seen, the operators $\hat{P}_{z\pm}$ as well as the operators $\hat{P}_{x\pm}$ form two different contexts: To be sure, $\hat{P}_{z+}\hat{P}_{z-}=\hat{P}_{z-}\hat{P}_{z+}$ while $\mathrm{ran}(\hat{P}_{z+}) \subseteq \mathrm{ker}(\hat{P}_{z-})$ and $\mathrm{ran}(\hat{P}_{z-}) \subseteq \mathrm{ker}(\hat{P}_{z+})$; analogously, $\hat{P}_{x+}\hat{P}_{x-}=\hat{P}_{x-}\hat{P}_{x+}$ as well as $\mathrm{ran}(\hat{P}_{x+}) \subseteq \mathrm{ker}(\hat{P}_{x-})$ and $\mathrm{ran}(\hat{P}_{x-}) \subseteq \mathrm{ker}(\hat{P}_{x+})$; but $\hat{P}_{z\pm}\hat{P}_{x\pm} \neq \hat{P}_{x\pm}\hat{P}_{z\pm}$. Let us represent these two contexts by the symbols $\mathcal{O}_z$ and $\mathcal{O}_x$ respectively.\\

\noindent The meet $\sqcap$ and join $\sqcup$ of the elements in $\mathcal{O}_x$ are given by\smallskip

\begin{equation} \label{25} 
    \hat{P}_{x+} \sqcap \hat{P}_{x-}
   \!
   =
   \hat{0}_2
   \;\;\;\;  ,
\end{equation}

\begin{equation} \label{26} 
    \hat{P}_{x+} \sqcup \hat{P}_{x-}
   \!
   =
   \hat{1}_2
   \;\;\;\;  ,
\end{equation}
\smallskip

\noindent where $\hat{0}_2$ and $\hat{1}_2$ are the zero and identity matrices respectively. Along these lines, the lattice operation $\underline{\sqcup}$ that corresponds to the exclusive disjunction $X_{x+} \,\underline{\lor}\, X_{x-}$ can be defined as\smallskip

\begin{equation} \label{27} 
    \hat{P}_{x+} \,\underline{\sqcup}\, \hat{P}_{x-}
    =
    \left(
       \hat{P}_{x+} \sqcup \hat{P}_{x-}
    \right)
    \sqcap
     \left(
       \hat{1}_2
       -
       \hat{P}_{x+} \sqcap \hat{P}_{x-}
    \right)
   =
   \hat{1}_2 (\hat{1}_2 -\hat{0}_2)
   =
   \hat{1}_2
   \;\;\;\;  .
\end{equation}
\smallskip

\noindent At this point, let us come back to the superposition (\ref{1}).\\

\noindent After the first experiment, the particle's spin state collapses into in the state $|\Psi_{z+}\rangle$ lying in the column space $\mathrm{ran}(\hat{P}_{z+})$ and the null space $\mathrm{ker}(\hat{P}_{z-})$ which -- under the valuation assumptions (\ref{12}) and (\ref{13}) -- implies that until the result of the second experiment is known, the propositions $Z_{+}$ and $Z_{-}$ are definite and have the truth-values 1 and 0 in that order. Explicitly,\smallskip

\begin{equation} \label{28} 
   |\Psi_{z+}\rangle
   =
   \left[
      \!\!
      \begin{array}{l}
         1\\
         0
      \end{array}
      \!\!
   \right]
   \in
   \bigg\{
   \left[
      \!\!
      \begin{array}{r}
         a\\
         0
      \end{array}
      \!\!
   \right]
   \bigg{|}\,
   a \in \mathbb{R}
   \bigg\}
   \iff
   \left\{
      \begin{array}{l}
         v(\hat{P}_{z+}) ={[\![ Z_{+} ]\!]}_v = 1\\
         v(\hat{P}_{z-}) ={[\![ Z_{-} ]\!]}_v = 0
      \end{array}
   \right.
   \;\;\;\;  .
\end{equation}
\smallskip

\noindent Then again, the state $|\Psi_{z+}\rangle$ does not lie in the column or null space of either projection operator in the nonselected context $\mathcal{O}_x$, namely,\smallskip

\begin{equation} \label{29} 
   |\Psi_{z+}\rangle
   =
   \left[
      \!\!
      \begin{array}{l}
         1\\
         0
      \end{array}
      \!\!
   \right]
   \notin
   \bigg\{
   \left[
      \!\!
      \begin{array}{r}
                a\\
         \pm a
      \end{array}
      \!\!
   \right]
   \bigg{|}\,
   a \in \mathbb{R}
   \bigg\}
   \;\;\;\;  .
\end{equation}
\smallskip

\noindent Based on the assumption (\ref{17}), this entails a truth-value gap for the propositions $X_{+}$ and $X_{-}$, explicitly,\smallskip

\begin{equation} \label{30} 
   |\Psi_{z+}\rangle
   \notin
   \left\{
      \begin{array}{l}
         \mathrm{ran}(\hat{P}_{x\pm})\\
         \mathrm{ker}(\hat{P}_{x\pm})
      \end{array}
   \right.
   \iff
   \left\{
      v(\hat{P}_{x\pm})
   \right\}
   =
   \left\{
      {[\![ X_{\pm} ]\!]}_v
   \right\}
   =
   \emptyset
   \;\;\;\;  ,
\end{equation}
\smallskip

\noindent even though the exclusive disjunction on the values of these propositions ought to be super-true:\smallskip

\begin{equation} \label{31} 
   v\left(
      \hat{P}_{x+} \,\underline{\sqcup}\, \hat{P}_{x-}
   \right)
   =
   v(\hat{1}_2)
   =
   {[\![ X_{x+} \,\underline{\lor}\, X_{x-} ]\!]}_v
   =
   1
   \;\;\;\;  .
\end{equation}
\smallskip

\noindent This gives the following logical account of the superposition (\ref{1}) compatible with quantum supervaluationism:\smallskip

\begin{equation} \label{32} 
   {[\![ Z_{+} ]\!]}_v
   =
   1
   =
   {[\![ X_{x+} \,\underline{\lor}\, X_{x-} ]\!]}_v
   \;\;\;\;  ,
\end{equation}

\begin{equation} \label{33} 
   \big\{
      {[\![ X_{\pm} ]\!]}_v
   \big\}
   =
   \emptyset
   \;\;\;\;  .
\end{equation}
\smallskip

\noindent According to this account, ahead of the result of the second experiment, the statement ``\textit{Out of two possible spins along the nonselected by the experiment axis, the particle possesses one or the other, but not both}'' is true even though the statement ``\textit{The particle possesses a spin along the nonselected by the experiment axis}'' has no truth value of any kind.\\

\section{Concluding remarks}  

\noindent Despite being non-bivalent, the presented above quantum supervaluationist account validates the principle of excluded middle. Indeed, in view of $\hat{P}_{x-} = \hat{1}_2 - \hat{P}_{x+}$, one can put $\neg X_{+}$ in the place of $X_{-}$ and get using (\ref{26}) that the disjunction $X_{+} \lor \neg X_{+}$ must be true (since it is super-true, that is, true in every admissible particle's spin state), namely,\smallskip

\begin{equation} \label{34} 
   v
   \left(
   \hat{P}_{x+} \sqcup (\hat{1} - \hat{P}_{x+})
   \right)
   =
   v(\hat{1}_2)
   =
   {[\![ X_{+} \!\lor \neg X_{+}]\!]}_v
   =
   1
   \;\;\;\;  .
\end{equation}
\smallskip

\noindent But together with that, the quantum supervaluationist account is, in general, \textit{not truth-functional}: E.g., the disjunction $X_{+} \!\lor \neg X_{+}$ is true although neither of its disjuncts, $X_{+}$ or $\neg X_{+}$, is true.\\

\noindent In addition, the conjunction $X_{+} \!\land \neg X_{+}$ exhibits the analogous non-classical feature: As stated by (\ref{25}), this connective must be false (since it is super-false, that is, false in every admissible particle’s spin state), namely,\smallskip

\begin{equation} \label{35} 
   v
   \left(
   \hat{P}_{x+} \sqcap (\hat{1} - \hat{P}_{x+})
   \right)
   =
   v(\hat{0}_2)
   =
   {[\![ X_{+} \land \neg X_{+}]\!]}_v
   =
   0
   \;\;\;\;  ,
\end{equation}
\smallskip

\noindent although it be that neither $X_{+}$ nor $\neg X_{+}$ is false.\\

\noindent According to the distributive law of classical logic, it must be that\smallskip

\begin{equation} \label{36} 
   Z_{+}
   \!\land
   \left(
      X_{+} \!\lor \neg X_{+}
   \right)
   =
   \left(
      Z_{+} \!\land X_{+}
   \right)
   \lor
   \left(
      Z_{+} \!\land \neg X_{+}
   \right)
   \;\;\;\;  .
\end{equation}
\smallskip

\noindent But, in the quantum supervaluationist account, even though the expression on the left side of the ``='' sign is associated with the truth (e.g., $\min\left\{ {[\![ Z_{+} ]\!]}_v, {[\![ X_{+} \!\lor \neg X_{+}]\!]}_v \right\} = 1$), the truth-functional interpretation of the expression on the right side of the ``='' sign is not possible, for, when $Z_{+}$ is definite, neither $X_{+}$ nor $\neg X_{+}$ has the truth-value (consequently, any function of ${[\![ Z_{+} ]\!]}_v$ and ${[\![ X_{\pm} ]\!]}_v$ would have no value). Thus, a statement that these two expressions are logically equivalent (i.e., have the same truth value) as well as a statement that these expressions are not logically equivalent has no meaning.
$\,$\footnote{\label{f3}It is worthy of note that such a conclusion is in line with \textit{the new quantum logic} proposed in \cite{Griffiths02,Griffiths14} in accordance with which binary connectives that join two sentences represented by the projection operators from different contexts have no meaning.\vspace{5pt}}\\

\noindent In that sense, the distributive law (\ref{36}) does not remain valid in the quantum supervaluationist account. Importantly however, this nonvalidness is not postulated but resulted from the supervaluation behavior of the propositions relating to the quantum superposition.\\

\section*{Acknowledgment}
\noindent The author is indebted to the anonymous referee for the incisive yet inspiring remarks helping him to improve the paper.\\

\end{document}